# The non-linear terahertz response of hot electrons in low-dimensional semiconductor superlattices: Suppression of the polar-optical phonon scattering

Anatoly A. Ignatov[1,2] [1)]

[1] *Institute for Physics of Microstructures, RAS, GSP-105, Nizhny Novgorod 603950, Russia*

[2] *University of Nizhny Novgorod, Nizhny Novgorod 603950, Russia*

……We study the response of low-dimensional semiconductor superlattices to strong terahertz fields on condition of a strong suppression of inelastic scattering processes of electrons caused by the polar-optical phonons. For our study we employ a balance equations approach which allows investigating the response of the superlattices to strong terahertz fields taking account of both the inelastic and the strongly pronounced elastic scattering of electrons. Our approach provides a way to analyze the influence of the Bloch dynamics of electrons in a superlattice miniband side by side with the effects of the electron heating on the magnitude and the frequency dependence of a superlattice current responsivity in the terahertz frequency band. Our study shows that the suppression of the inelastic scattering caused either by a reduction of the superlattice dimensionality by lateral quantization or by a strong magnetic field application can give rise to a huge enhancement of the current responsivity. This enhancement can be interpreted in terms of the well pronounced electronic bolometric effect occurring due to the efficient electron heating in the low-dimensional superlattices by the incident terahertz fields.

---

[1)] Author to whom correspondence should be addressed. Electronic mail: ign@ipm.sci-nnov.ru

## I. INTRODUCTION

In recent years, semiconductor superlattices have received considerable attention from the viewpoint of the creation of new detectors[1-5], oscillators[6-10] and frequency converters[11-13] of the terahertz and sub terahertz radiation. In this regard, the study of new transport phenomena associated with the Bloch electron dynamics and the electron heating in superlattices appears to be very interesting. Earlier, significant attention was paid to the study of the influence of the electron scattering on the high-frequency conductivity of superlattices. In particular, it has been shown that the heating processes formed both by the elastic and the inelastic scattering of electrons can substantially determine the frequency region of the negative differential conductivity (NDC) effect in this structures[14-16].

On the other hand, semiconductor superlattices have been employed for the study of such fascinating effects as the emission of Bloch oscillations of electrons in strong dc fields both at low and room temperatures[17-19], the inverse Bloch oscillator effect[20], the dynamic localization of electrons in novel super-radiance lattices[21], the chaotic electron diffusion through stochastic webs in superlattices subjected to strong tilted magnetic fields[22,23], and many other ones (for a review see also Refs. 24-26). One can also note that the superlattice concept is a clue component in novel device constructions, such as quantum cascade lasers[27].

In studies of the superlattice high-frequency response to strong terahertz fields the photon-assistant tunneling model is the most commonly used. In terms of the semi-classical miniband transport the basic equations of this model can be derived[28-31] making use of the original Esaki-Tsu path integral solution of the Boltzmann equation[32, 33]. In this case it is supposed that the collision integral in the Boltzmann equation can be presented in an approximation of the single relaxation time describing in fact only the processes of inelastic scattering of electrons. In the static case this approach results in the dc Esaki-Tsu current-voltage curve of the superlattice which is widely used in simulations of the high-frequency response of the superlattices as well as for interpreting of a number of experimental data[20-26].

However, a number of theoretical[14] and experimental[7, 34,] papers devoted to the study of both the static current-voltage curves of the superlattices and the terahertz field response of these structures suggest that the inelastic scattering of electrons should be considered in combination with the elastic scattering. In particular, X-ray characterization of the superlattices indicate that the interface roughness of about one monolayer thickness may be responsible for a dramatic reduction of the electron current along the superlattice axis[35] due to the presence of the elastic scattering of electrons. Moreover, the observation of the quasi-static and the dynamic interaction of the strong terahertz fields with miniband electrons in semiconductor superlattices[34] were observed in the superlattice samples where the elastic scattering was prevailing. Therefore, investigation of the high frequency response of the superlattices beyond the scope of the photon-assistant tunneling theory seems to be very relevant.

A special impact on the ac/dc transport properties of the superlattices the elastic scattering of electrons can have in the case of reducing the superlattice dimensionality as a result of either the electron lateral quantum confinement[36-38] or in the case of the strong quantizing magnetic field application[39-41]. In this case both the energy spectrum of electrons and the electron scattering processes can be essentially modified. Consequently, the basic parameters describing the interaction of the terahertz fields with the superlattices such as the terahertz conductance and the current responsivity can be strongly influenced by the presence of the elastic scattering.

Recently, quantum nanostructures such as quantum wires and coupled quantum dots have attracted much attention because of the possibility to tailor their electronic and transport properties (for a review see Ref. 42). Among others, one can mention the one-dimensional hetero-structures in semiconductor nanowhiskers[43] and InGaAs/GaAs quantum-dot chains[44]. We would also mention the synthesis of extended graphdiyne wires that could allow one to investigate the non-linear electronic transport related to Bloch oscillations[45,46], the proposal to realize the spin-dependent terahertz oscillator based on hybrid graphene superlattices[47], the possibility of enhancing of spin injection in a spin-filter superlattice[48], theoretical investigation of the transport properties of superlattice nanowires for thermoelectric applications[48], and proposal for the third harmonic generation in a terahertz driven modulated nanovires[50,51].

The transition of electron motion in a superlattice to a lower dimensionality can lead to a significant change in the nature of electron scattering. In particular, it has been demonstrated[36] that if the quasi-1D superlattice miniband width is less than the polar-optical phonon (POP) frequency the scattering of electrons on polar-optical phonons can be essentially suppressed. Due to the fact that the POP scattering is the main mechanism governing the energy dissipation in semiconductors, this circumstance can give rise to a strong change of the current flow in the superlattices. One can also expect that the POP scattering suppression may lead to a substantial change of the high-frequency response of the superlattices in the terahertz frequency band.

An alternative way to create controllable quasi-one-dimensional energy minibands in a superlattice is the application of the strong (quantizing) magnetic fields oriented along the superlattice axis[39-41]. Several works demonstrated that the current flow through the superlattice can be hugely suppressed when the electron motion perpendicular to the superlattice axis is strongly confined by the quantizing magnetic field[52, 53]. These studies experimentally demonstrated that the reduced dimensionality restricts the range of inelastic scattering processes for the conduction electrons that gives rise to an essential increase of the inelastic scattering times and to a corresponding dramatic decrease of the electrical current.

The present paper is devoted to a theoretical investigation of detection of the terahertz photons by the low-dimensional semiconductor superlattices. Our studies are carried out using the approach going beyond the photon-assisted tunneling

4theory. A special emphasis is paid to the superlattices placed into strong (quantizing) magnetic fields leading to suppression of the electric current[52-53] due to the elimination of the POP scattering. We are based on the balance equations approach which allows one to take into account both elastic and inelastic scattering of electrons in a superlattice miniband. Earlier, a similar balance equations approach has been used for discussion of the anatomy of the transferred-electron effect in III-V semiconductors[55], for the analyses of the terahertz response of hot electrons in dilute nitride Ga(AsN) alloys[56,57], and for investigation of the high-field transport and terahertz generation in GaN[58]. We believe that the results obtained in the present paper will be useful for development of novel ultrafast and sensitive detectors based on superlattices with suppressed POP scattering of electrons operating in the terahertz frequency band.

The paper is organized as follows. In Sec. II we discuss the specifics of the energy spectrum of electrons, caused either by the lateral confinement of their motion across the axis of the superlattice, or by the application of the strong (quantizing) magnetic field applied along the superlattice axis. We introduce the balance equations for the study of the ac/dc transport in the low-dimensional superlattices taking account of both the elastic and inelastic scattering mechanisms side by side with the Bloch dynamics of electrons in the superlattice miniband. We emphasize that the change of sign of the effective mass of electrons caused by their heating in strong ac/dc fields may have a dramatic impact on both the static and the high frequency response of the superlattices.

In Sec. III, we study the response of electrons in the low-dimensional superlattices to strong ac/dc fields beyond the photon-assisted tunneling theory, i.e. in the case when the elastic scattering of electrons has a significant impact on electron transport phenomena. We derive the time-dependent transport equation for the electron current in a superlattice that takes into account i) Bloch oscillations of electrons in strong dc fields, ii) parametric effects in the superlattice caused by the strong ac fields, and iii) the presence of the strong elastic scattering of electrons giving rise to the Bloch oscillations damping.

Then, we analyze the influence of the strong elastic scattering on the dc current-voltage curves of the superlattices. We emphasize that the strong elastic scattering has a critical impact on the dc current-voltage curves of the superlattices leading to its fundamental difference from the Esaki-Tsu dc current-voltage curve. Specifically, we show that the suppression of the POP inelastic scattering gives rise to a dramatic decrease of both the peak voltage and the peak current of the dc current-voltage curves.

Afterwards, we calculate the ac conductance and the dc current change in the superlattice irradiated with the strong terahertz fields. We show that both of these values can be strongly different from predictions of the photon-assistant tunneling theory in specific conditions of the inelastic POP scattering suppression. We present the result of the self-consistent calculation of a superlattice current responsivity taking into account the resonant properties of the system caused by the presence of the polar-optical phonons. Based on the analytical equations obtained in this section we derive the expression for the characteristic frequency-dependent dc voltage at which the transition between terahertz loss and gain in the superlattice takes place in the presence of the strong elastic scattering. We find that the transition from loss to gain in the superlattice corresponds to the change of the sign of the superlattice current responsivity, i.e. to the occurrence of the positive photocurrent in the superlattice response.

In Sec. IV we present and discuss the results obtained on the basis of the analytical approach developed in the previous sections. When describing the nonlinear response of the superlattices we focus our attention on the experimental situation, when the suppression of the inelastic polar-optical scattering was caused by the application of the strong magnetic fields parallel to the axis of the superlattice. In this case the role of the elastic scattering in the dc current-voltage curve as well as in the high frequency response of the superlattice becomes a crucial point.

Firstly, we examine the effect of suppression of both the static peak current and the static peak voltage in the superlattice when the strong magnetic fields are applied along the superlattice axis. We show that the dependence of the static peak current on the inelastic scattering times can be found in an excellent agreement with the experimental data presented in Refs. 52 and 53.

Secondly, we explore the ultimate frequency of the terahertz gain in the superlattices in the presence of the strong elastic scattering. We demonstrate that the presence of the elastic scattering of electrons results in a substantial reduction of the ultimate frequency of the terahertz gain compared to the frequency of the Bloch oscillations for the same dc electric field values. We show that this fact can have a significant impact on the value of the high-frequency fields produced in the superlattices by external sources of electromagnetic radiation.

......Thirdly, we examine the frequency dependence of the current responsivity of the superlattice with different lateral sizes in the absence of magnetic fields. By way of example, calculations are made for the superlattices experimentally investigated in Refs. 52 and 53. We demonstrate that the decrease in the superlattice lateral size can lead to a drastic enhancement of the current responsivity. We analyze the frequency dependence of the current responsivity at different lateral sizes of the superlattices for optimization of the superlattice-based detectors in the terahertz frequency band. Then, we discuss the existence of the strong resonant (anti-resonant) response of the superlattice-based detectors at frequencies of longitudinal (transversal) optical phonons.

Finally, in the last part of Sec. IV we discuss the influence of the strong magnetic field parallel to the superlattice axis on the absolute value of the current responsivity in the terahertz frequency band at different values of the applied dc voltage. We emphasize that at certain values of the dc voltage in the negative differential conductivity region the huge dips can occur corresponding to the transition from the absorption to gain of the terahertz fields in the superlattice. On the other hand, we show that at certain values of the dc voltage applied to the superlattice in the positive negative conductivity region, the current responsivity can dramatically increase with the application of the strong magnetic fields. We argue that this phenomenon can arise due to an increase in the efficiency of electron heating in the superlattices caused by suppression of the POP scattering. We believe that in this regard, the low-dimensional semiconductor superlattices could be interesting for development of the new types of the high-speed sensitive detectors operating in the terahertz frequency band.

In Sec. V, the conclusions are briefly summarized.

## II. DESCRIPTION OF THE MODEL

In a quasi-1D superlattice electrons are spatially confined in directions perpendicular to the superlattice axis $z$ along which an artificial periodic potential $U(z)$ is created (see Fig. 1). In particular, for systems consisting of the coupled quantum boxes (quantum wire superlattices) with the $D \times D$ cross-section aria the electron energy can be written as[36]

$$\varepsilon(p_z, k, p) = \varepsilon_{SL}(p_z) + \varepsilon_{k,p}, \qquad (1)$$

where $\varepsilon_{SL}(p_z)$ is the energy dispersion along the superlattice axis $z$, $p_z$ is the electron momentum along the superlattice axis, $\varepsilon_{k,p} = (\pi\hbar)^2(k^2 + p^2)/2mD^2$ is the quantized energy in the quantum-wire cross-section ($xy$ plane), $k, p = 1, 2, ...$, are the integers, and $m$ is the electron effective mass of electrons in the $xy$ plane. In this case the length $D$ plays a role of the characteristic electron localization length in the transversal $xy$ plane.

On the other hand, the application of a strong magnetic field $B$ along the 3D superlattice axis $z$ (see Fig. 1) represents an alternative way to create the quasi-1 D superlattice system. In this case, the electronic motion in the $xy$ plane is also quantized and the electron energy can be written as[39]

$$\varepsilon(p_z, k) = \varepsilon_{SL}(p_z) + \left(k + \frac{1}{2}\right)\hbar\omega_c, \qquad (2)$$

where $\varepsilon_{SL}(p_z)$ is the energy dispersion of the miniband in direction parallel to the superlattice axis $z$, $\omega_c = eB/m$ is the cyclotron frequency, $m$ is the effective mass in the $xy$ plane, and $k = 0, 1, 2...$ is the integer. The characteristic length of electron localization in the $xy$ plane in this case can be estimated as $l_c = \sqrt{\hbar/(m\omega_c)}$.

In both cases the miniband structure can be approximated by cosine shape in a tight-binding approximation[36-39], i.e.

$$\varepsilon_{SL}(p_z) = \frac{\Delta}{2}\left[1 - \cos\left(\frac{p_z d}{\hbar}\right)\right], \qquad (3)$$

where $\Delta$ is the superlattice miniband width, and $d$ is the superlattice period.

If the characteristic quantized energies are high enough in comparison with the miniband width $\Delta$, i.e. if

$$\Delta\varepsilon_{1,2} > \Delta, \quad \hbar\omega_c > \Delta, \qquad (4)$$

where $\Delta\varepsilon_{1,2} = \varepsilon_{1,2} - \varepsilon_{1,1} = \varepsilon_{2,1} - \varepsilon_{1,1}$, the gap between the ground and the first quantized sub-bands emerges (see Fig. 1). This circumstance gives rise to a quasi one-dimensional character of the conductance electron motion in the superlattice. Moreover, for certain values of the quasi-1D superlattice parameters it becomes possible to suppress the polar-optical phonon (POP) scattering. Namely, if the miniband width $\Delta$ is smaller than the POP energy $\hbar\omega_{POP}$, i.e. if

$$\Delta < \hbar\omega_{POP} \qquad (5)$$

it is possible to suppress POP emission and absorption which is responsible for the dominant energy dissipation process in polar semiconductors[36-39]. In a phenomenological way, this effect can be taken into account by the reduction of the inelastic scattering relaxation rate which is involved in the picture of the balance-equation description of the superlattice electron transport[52, 53].

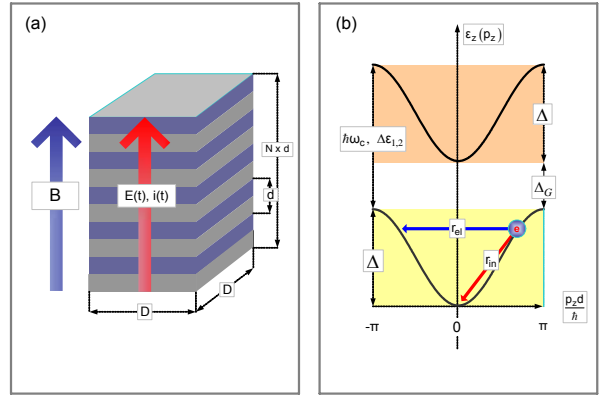

FIG. 1. (a) Schematic presentation of the superlattice structure placed in a strong magnetic field $B$ under irradiation by the terahertz fields. (b) Quantum size confinement minibands or Landau level minibands giving rise to the quasi one-dimensional motion of electrons in a superlattice.

For an analysis of the nonlinear response of our superlattice to intense THz fields, we use the solution of the balance equations for the electron time-dependent mean energy $\bar{\varepsilon}$ and the time-dependent mean electron velocity $\bar{v}$. In a one-miniband approximation these equations can be presented as[14, 28, 31]

$$\dot{\bar{v}} = \mu(\bar{\varepsilon})E(t)r_v - r_v\bar{v}, \tag{6}$$

$$\dot{\bar{\varepsilon}} = eE(t)\bar{v} - r_{in}(\bar{\varepsilon} - \bar{\varepsilon}_0), \tag{7}$$

where $\bar{v}$ is the mean velocity of electrons in the superlattice miniband, $\bar{\varepsilon}$ is the mean energy of electrons, $E(t)$ is the time-dependent electric field directed along the superlattice axis $z$, $\mu(\bar{\varepsilon}) = \mu_0(1 - 2\bar{\varepsilon}/\Delta)$ is the mobility of electrons depending on the mean electron energy $\bar{\varepsilon}$, $\mu_0 = e/r_v m_0$ is the low-field mobility of electrons, $m_0 = 2\hbar^2/\Delta d^2$ is the effective mass of electrons in the miniband bottom in the direction of the superlative axis $z$,

$$r_v = r_{in} + r_{el} \tag{8}$$

is the mean velocity relaxation rate, which includes the energy (inelastic) relaxation rate $r_{in}$ and the elastic relaxation rate $r_{el}$, $d$ is the superlattice period, $\hbar$ is the Planck's constant, $e$ is the elementary charge, and $\bar{\varepsilon}_0$ is the mean electron energy in thermal equilibrium. We will also use the following definitions for the inelastic $\tau_{in} = 1/r_{in}$ and elastic $\tau_{el} = 1/r_{el}$ scattering relaxation times, respectively.

Equation (6) describes the time evolution of the mean electron velocity $\bar{v}(t)$ taking account of the dependence of the electrons mobility on the mean energy $\bar{\varepsilon}$, while Eq. (7) represents the energy conservation low describing the relaxation of the non-equilibrium mean energy $\bar{\varepsilon}$ to its equilibrium value $\bar{\varepsilon}_0$ at $E(t) \to 0$. In the case when $kT, \varepsilon_F \ll \Delta$, where $kT$ is the equilibrium thermal excitation energy, $\varepsilon_F = \hbar^2(3\pi^2 n)^{2/3}/(2m_{eff})$ is the Fermi energy of degenerate electrons, $m_{eff} = m_0^{1/3} m^{2/3}$ is the density of states effective mass near the miniband bottom, $m$ is the electron effective mass perpendicular to the layers, and $n$ is the conductance electron density in the superlattice miniband, one can assume $\bar{\varepsilon}_0 \to 0$.

We can also note that the Eq. (6) can be regarded as a "Newton's equation" for the mean velocity of electrons $\bar{v}$ where the mean electrons effective mass $m(\bar{\varepsilon}) = m_0/(1 - 2\bar{\varepsilon}/\Delta)$ depends on the mean electrons energy $\bar{\varepsilon}$. If the mean electrons energy $\bar{\varepsilon}$ exceeds the half of the miniband width $\Delta$, the mean effective mass $m(\bar{\varepsilon})$ becomes negative[31]. The right hand terms in Eqs. (6) and (7) are responsible for electron heating effect in the superlattice, while the left hand terms take into account the time evolution of the electron mean velocity $\bar{v}(t)$ and the mean energy $\bar{\varepsilon}(t)$.

The negative effective mass in the superlattice minibands leads to a dramatic change of the current flow in superlattices in both static and high frequency electric fields in comparison with the bulk semiconductors giving rise to the negative differential conductance (NDC) effect. Some basic physical principals of the gain of the terahertz fields due to the negative electron masses have been recently discussed in Refs. 58 and 59.

Applicability of Eqs. (6) and (7) for the description of the non-linear electromagnetic response of the superlattices has been confirmed by a number of papers comparing the results of the analytical calculations based on the balance equations with the results of the Monte-Carlo modeling[60,61]. In particular, Ref. 62 demonstrates that calculations of the small ac field conductivity of the superlattice based on Eqs. (6) and (7) are in excellent agreement with the Monte-Carlo calculations, whereas the single-relaxation-time approximation ignoring the elastic scattering systematically overestimates the semi-classical gain occurring due to the negative differential conductance in superlattices.

## III. RESPONSE OF A SUPERLATTICE TO STRONG AC/DC FIELDS

This broad section deals with the analytical solution of Eqs. (6) and (7) in order to study the response of the superlattice to strong ac/dc fields. In Sec. IIIA the dynamical Bloch oscillations equation which allows one to apply the perturbation technique for the calculation of the superlattice current responsivity is derived. Section IIIB discusses the dc current-voltage curves of the superlattice in the presence of the strong elastic scattering of electrons. Section IIIC presents the results of the calculation of the ac conductance of the superlattice as well as the terahertz field induced dc current change in the superlattice. Section IIID treats the one of the main characteristics of the superlattice non-linear response to the strong terahertz fields, i.e. the superlattice current responsivity. Finally, Sec. IIIE reveals the intimate relationship between some features of the frequency dependence of the superlattice current responsivity and the frequency region of the terahertz gain.

### A. Dynamical Bloch oscillation equation

In order to analyze the influence of the strong elastic scattering on the terahertz superlattice response we reduce Eqs. (6) and (7) to the unique equation for the normalized time-dependent conductance electrons current $I(t)$

$$\ddot{I}(t) + \left[(r_v + r_{in}) - \frac{\dot{\Omega}(t)}{\Omega(t)}\right] \times \dot{I}(t) + \left[\Omega(t)^2 + r_{in} r_v - r_v \times \frac{\dot{\Omega}(t)}{\Omega(t)}\right] \times I(t) = 2(r_v r_{in})^{1/2} \times \Omega(t), \qquad (9)$$

where $\Omega(t) = eE(t)d/\hbar = eV(t)/N\hbar$ is the dynamical Bloch frequency, $V(t)$ is the total voltage applied to the superlattice, $N$ is the total number of the superlattice periods, $I(t) = i(t)/i_p$ is the normalized time-dependent conductance electrons current, $i(t) = enS\bar{v}[E(t)]$ is the conductance electrons current throw the superlattice, $\bar{v}[E(t)]$ is the time-dependent electrons mean (drift) velocity, $E(t)$ is the time-dependent electric field, $i_p = enSv_p$ is the characteristic (peak) conductance electrons current in the superlattice, $S = D^2$ is the superlattice cross-section aria, $v_p = v_0 \left[ r_{in} / (r_{in} + r_{el}) \right]^{1/2} / 2$ is the characteristic (peak) electrons mean velocity, $v_0 = \Delta d / 2\hbar$ is the maximum value of the electrons velocity in the superlattice miniband, and $\delta = r_{in} / r_v = r_{in} / (r_{in} + r_{el})$ is the ratio of the relaxation rates.

If the constant electric field $E_0 = const$ is suddenly applied to the superlattice Eq. (9) reduces to the equation describing the time evolution of the damped Bloch oscillations of electrons[14]

$$\ddot{I} + (v_v + v_{in})\dot{I} + (v_v v_{in} + \Omega_B^2)I = 2(v_v v_{in})^{1/2} \times \Omega_B, \tag{10}$$

where $\Omega_B = eE_0 d / \hbar$ is the Bloch frequency of electrons in the static electric field $E_0$.

Equation (10) describes the transient processes in semiconductor superlattices. By means of this equation it was found a certain criteria determining connection between the occurrences of Bloch oscillations in the transient processes and the ac negative differential conductance effect[14]. Meanwhile, the general Eq. (9) represents the linear equation of second order with the time dependent coefficients and the time dependent external force. Consequently, it can justify the occurrence of a number of parametric effects in superlattices irradiated by intense high-frequency fields[63, 64].

In the present paper, for the self-consistent analysis of the nonlinear superlattice response to strong terahertz fields, we employ the perturbation technique. We suppose that the dc voltage $V_0$ and, in addition, the small alternating sinusoidal voltage $V_1(t)$ is applied to the superlattice, i.e.

$$V(t) = V_0 + \left[ \frac{V_1}{2} \exp(i\omega t) + c.c. \right], \tag{11}$$

where $V_1$ is the small complex amplitude of the ac voltage, and $\omega$ is its frequency.

In this case solution of Eq (9) for the time-dependent conduction electrons current $i(t)$ in the superlattice can be presented as

$$i(t) = i_0(V_0) + \left[ G(\omega, V_0) \times \frac{V_1}{2} \times \exp(i\omega t) + c.c. \right] + \bar{\delta} i_0(\omega, V_0), \tag{12}$$

where $i_0(V_0)$ is the dc (static) component of the current, $G(\omega, V_0)$ is the complex ac conductance of the superlattice, and $\overline{\delta i_0}(\omega, V_0)$ is the terahertz field induced dc current change.

## B. DC current-voltage curve

Using Eqs. (6) and (7), for the dc component of the current $i_0(F_0)$ and for the mean electron energy $\overline{\varepsilon}_0(F_0)$ we obtain

$$i_0(F_0) = i_p \times \frac{2F_0}{1+F_0^2}, \tag{13}$$

$$\overline{\varepsilon}_0(F_0) = \frac{\Delta}{2} \times \frac{F_0^2}{1+F_0^2}, \tag{14}$$

where $F_0 = V_0/V_p = E_0/E_p$ is the normalized dc voltage (electric field) in the superlattice, $E_0 = V_0/L$ is the dc electric field strength, $V_p = L \times E_p = N \times \hbar \left[ r_{in}(r_{in} + r_{el}) \right]^{1/2} / e$ is the characteristic (peak) voltage, $L = N \times d$ is the total superlattice length, $E_p = \hbar (r_{in} r_v)^{1/2} / ed = \hbar \left[ r_{in}(r_{in} + r_{el}) \right]^{1/2} / ed$ is the characteristic (peak) dc electric field, $i_p = j_p \times S$ is the peak dc current, $j_p = env_p$ is the peak current density, $j(F_0) = env_d(F_0)$ is the dc current density, $v_d(F_0) = v_p \times 2F_0/(1+F_0^2)$ is the dc drift velocity of electrons given as function of the normalized dc electric field $F_0$, and $v_p = (\Delta d / 4\hbar) \times (r_{in}/r_v)^{1/2} = (\Delta d / 4\hbar) \times \left[ r_{in}/(r_{in} + r_{el}) \right]^{1/2}$ is the dc peak drift velocity.

As it is seen from Eq. (13) that at $V_0 = V_p$ or, equivalently, at $E_0 = E_p$ the dc current in the superlattice reaches its maximum (peak) value $i_p$ or, equivalently, the electrons drift velocity $v_d(F_0)$ reaches the maximum (peak) drift velocity value $v_p$. At higher voltage (dc field) values the dc current-voltage curve demonstrates the negative differential conductance (NDC) effect, i.e. $di_0(V_0)/dV_0 < 0$ or $dv_d(F_0)/dF_0 < 0$ at $V_0 > V_p$ or at $E_0 > E_p$.

On the other hand, according to Eq. (14) due to the effect of electron heating the mean energy $\overline{\varepsilon}_0(F_0)$ is essentially changing with changing of the applied dc field values $F_0$. This circumstance shows that the occurrence of the negative differential conduction in superlattices is inherently linked with the electron heating. It is important to note that according to Eq. (7), the value of the mean velocity of electrons in a stationary case $v_d(E_0)$ is proportional to the inelastic relaxation rate $r_{in}$, i.e. $v_d(E_0) = r_{in}\bar{\varepsilon}_0(E_0)/eE_0$. Thus, suppression of the polar-optical scattering can result in a substantial change of the electron conduction in the superlattice.

In a particular case, when the elastic scattering is neglected, i.e. when $r_{el} \to 0$, Eq. (13) results in the Esaki-Tsu dc current-voltage curve, which can be presented as[32, 33]

$$i_{ET}(\Omega_B) = 2i_{pET} \times \frac{\Omega_B \tau_{in}}{1 + (\Omega_B \tau_{in})^2}, \tag{15}$$

where $i_{pET} = j_{pET} \times S$ is the Esaki-Tsu peak current, $j_{pET} = env_{pET}$ is the Esaki-Tsu peak current density, and

$$v_{pET} = \frac{\Delta d}{4\hbar} \tag{16}$$

is the Esaki-Tsu peak drift velocity.

In terms of the Esaki-Tsu equations for the peak current $i_{pET} = S \times en\Delta d/4\hbar$ and the peak velocity $V_{pET} = N \times \hbar r_{in}/e$ the calculated peak current $i_p$ and the peak voltage $V_p$ in the presence of elastic scattering can be written presented as

$$i_p = i_{pET} \times \left(\frac{r_{in}}{r_{in} + r_{el}}\right)^{1/2} = S \times \frac{en\Delta d}{4\hbar} \times \left(\frac{r_{in}}{r_{in} + r_{el}}\right)^{1/2}, \tag{17}$$

$$V_p = V_{pET} \times \left(\frac{r_{in} + r_{el}}{r_{in}}\right)^{1/2} = N \times \frac{\hbar}{e} \times r_{in}^{1/2}(r_{in} + r_{el})^{1/2}. \tag{18}$$

In small dc electric fields $E_0 \ll E_p$ for the dc drift velocity in the superlattice $v_d(E_0)$ Eq. (13) yields

$$v_d(E_0) = \mu_0 E_0, \tag{19}$$

where

$$\mu_0 = \frac{e\tau_v}{m_0} = \frac{e}{m_0(r_{in} + r_{el})} \tag{20}$$

is the low-field electron mobility, and $\tau_v = 1/(r_{in} + r_v)$ is the mean electron velocity relaxation time.

Note that in the condition of the inelastic scattering suppression, i.e. if $r_{in} \ll r_{el}$, we obtain $\tau_v \to 1/r_{el}$. In this case the electron low-field mobility $\mu_0$ is determined only by the elastic scattering mechanisms, i.e. $\mu_0 \to (e/m_0)\tau_{el}$. On the contrary, for the high values of the dc electric fields $E_0 \gg E_p$ Eqs. (13) and (14) yield

$$\bar{\varepsilon}_0(E_0) \to \frac{\Delta}{2}, \quad (21)$$

$$v_d(E_0) \to v_0 \frac{\hbar r_{in}}{eE_0 d}. \quad (22)$$

In this case the mean electron energy $\bar{\varepsilon}_0(E_0)$ saturates due to the heating of electrons by the dc electric field and the value of the dc drift velocity $v_d(E_0)$ depends only on the electrons inelastic relaxation rate $r_{in}$.

Qualitatively, the physical significance of Eq. (13) is easy to understand if one accepts the idea that in the approximation of the effective relaxation times the dc drift velocity obeys to the generalized Matthiessen's rule, i.e. $1/v_d = 1/v_d^{lf} + 1/v_d^{hf}$, where $v_d^{lf} = \mu_0 E_0$ is the low-field drift velocity, $v_d^{hf} = P_{loss}/eE_0$ is the high-field limit of the drift velocity, and $P_{loss} = \Delta/2\tau_{in}$ is the loss power per particle of the heated electrons[65].

It is important to note that Eqs. (17) and (18) allow one to estimate the inelastic and elastic scattering rates $r_{in}$ and $r_{el}$ on the basis of the data obtained from the measurements of the dc peak current $i_p$ and the dc peak voltage $V_p$. Note that both the peak current $i_p \propto [\tau_{el}/(\tau_{in}+\tau_{el})]^{1/2}$ and the peak voltage $V_p \propto \tau_{in}^{-1}[(\tau_{in}+\tau_{el})/\tau_{el}]^{1/2}$ vanish if $\tau_{in} \to \infty$. Therefore, it can be concluded that there is no current flow in the superlattice in the absence of the inelastic scattering of electrons. This circumstance demonstrates the fundamental link between the current flow and the energy dissipation in superlattices[52, 53].

## C. AC conductance and dc current change

Within the framework of the perturbation technique, the following expressions for the terahertz field induced dc current change $\bar{\delta}i_0(\omega,V_0)$ and for the complex ac conductance of the superlattice $G(\omega,V_0)$ can be obtained from Eq. (9) and (12):

$$\bar{\delta}i_0(\omega,V_0) = \frac{1}{4}|V_1|^2 \times \bar{i}_0''(\omega,V_0), \quad (23)$$

$$G(\omega,V_0) = G_0 \times \frac{1}{1+F_0^2} \times \frac{(1+i\omega\tau_{in}) - F_0^2}{(1+i\omega\tau_{in})(1+i\omega\tau_{in}\delta) + F_0^2}, \quad (24)$$

where

$$\bar{i}_0''(\omega,V_0) = -2\frac{i_p}{V_p^2} \times I(F_0) \times \mathrm{Re}\left\{\frac{G(\omega,V_0)}{G_0} + \frac{1}{1+i\omega\tau_{in}}\left[\frac{G(\omega,V_0)}{G_0} + \frac{1}{1+F_0^2}\right]\right\} \quad (25)$$

14is the generalized second derivative of the dc current-voltage curve taken at the finite frequency $\omega$, $G_0 = 2i_p / V_p = \sigma_0 S / L$ is the conductance of the superlattice at $\omega, F_0 \to 0$, $\sigma_0 = 2j_p / E_p$ is the low-frequency/field conductivity, $\tau_{in} = r_{in}^{-1}$ is the inelastic scattering relaxation time, $\delta = r_{in} / r_v = r_{in} / (r_{in} + r_{el})$ is the ratio of the relaxation rates, and $I(F_0) = 2F_0 / (1 + F_0^2)$ is the normalized dc current-voltage curve of the superlattice.

Equation (24) for the complex conductance $G(\omega, V_0)$ describes the processes of absorption and gain of the terahertz fields in superlattices placed in a strong dc electric field in the presence of the elastic scattering. At the same time, Eq. (25) for the generalized second derivative of the dc current $\bar{i}_0''(\omega, V_0)$ is responsible for detection of the terahertz photons. It is important to note that the value of $\bar{i}_0''(\omega, V_0)$ is essentially determined by the real part of the conductance $G(\omega, V_0)$. This circumstance establishes the principal relationship between the process of the absorption or emission of photons, and the process of their detection.

In the particular case when the elastic scattering of electrons is completely neglected, i.e. when $r_{el} \to 0$, $\delta \to 1$, Eqs. (24) and (25) deliver the real part of the ac conductance $\operatorname{Re} G(\omega, V_0)$ and the generalized second derivative $\bar{i}_0''$ in the finite difference form, i.e.

$$\operatorname{Re} G_{ET}(\omega, V_0) = \frac{i_{ET}(V_0 + N\hbar\omega/e) - i_{ET}(V_0 - N\hbar\omega/e)}{2(N\hbar\omega/e)}, \tag{26}$$

$$\bar{i}_{0ET}''(\omega, V_0) = \frac{i_{ET}(V_0 + N\hbar\omega/e) - 2i_{ET}(V_0) + i_{ET}(V_0 - N\hbar\omega/e)}{(N\hbar\omega/e)^2}. \tag{27}$$

Equations (26) and (27) describe the high-frequency response of the superlattice within the framework of the photon-assistant tunneling theory that was originally applied for the analysis of the non-linear superlattice response in the THz frequency band[28-31].

It is important to note that Eq. (27) can be obtained taking the limit of the small values of the incident ac voltage amplitude $V_\omega \to 0$ from the general equation describing the time-averaged component of the dc current-voltage curve of the terahertz field irradiated superlattice[28-31]

$$\bar{i}_{dc}(\Omega_B) = \sum_{n=-\infty}^{\infty} J_n\left(\frac{eE_\omega d}{\hbar\omega}\right) \times i_{ET}(\Omega_B + n\hbar\omega), \tag{29}$$

where $\bar{i}_{dc}(\Omega_B) = (1/T)\int_T i(t)dt$, $T = 2\pi/\omega$ is the period of the incident ac field, $E_\omega = V_\omega/Nd$ is the ac field amplitude inside the superlattice, $J_n(x)$ is the Bessel function of the $n^{th}$ order, $n$ is the integer, and $i(t)$ is the conduction electrons current throw the superlattice that can be found analytically at $r_{el} \to 0$, $\delta \to 1$ as a solution of Eqs. (6) and (7) (for discussion of the analytical solution for $i(t)$ in this case see Ref. 31).

Meanwhile, the presence of the elastic scattering of electrons $r_{el} \neq 0$, $\delta < 1$ can have a strong influence on both the real part of the superlattice conductance $\text{Re}\,G(\omega, V_0)$ responsible for absorption (emission) of the terahertz fields as well as on the generalized second derivative $\bar{i}_0''(\omega, V_0)$ which is responsible for the terahertz photons detection. Equations (24) and (25) derived for these quantities can naturally combine the effects related to the Bloch dynamics of electrons in the superlattice miniband, the effects caused by the relaxation of the mean electrons energy induced by the inelastic scattering, and the effects of the electrons motion chaotization due to elastic scattering.

At small frequencies of the applied ac fields $\omega\tau_{in} \ll 1$ Eqs. (24) and (25) reduce to the first and second ordinary derivatives of the superlattice dc current-voltage curve, i.e.

$$G(0, V_0) \to \frac{di_0(V_0)}{dV_0} = 2\frac{i_p}{V_p}\frac{(1-F_0^2)}{(1+F_0^2)^2}, \qquad (29)$$

$$\bar{i}_0''(0, V_0) \to \frac{d^2 i_0(V_0)}{dV_0^2} = -4\frac{i_p}{V_p^2}\frac{F_0(3-F_0^2)}{(1+F_0^2)^3}. \qquad (30)$$

It is important to note that at small frequencies $\omega\tau_{in} \ll 1$ for the dc superlattice conductance we obtain $G(0, F_0) \propto i_p/V_p \propto 1/(r_{in} + r_{el})$. When $r_{in} \ll r_{el}$, the dc superlattice conductance $G(0, F_0)$ does not depend on the inelastic scattering rate $r_{in}$. In this case it is determined solely by the elastic scattering relaxation rate $r_{el}$. In contrast to this, the second derivative of the dc current throw the superlattice $\bar{i}_0''(0, F_0) \propto i_p/V_p^2 \propto 1/r_{in}^{1/2}(r_{in} + r_{el})^{3/2}$ rapidly increases with decreasing of the inelastic scattering rate $r_{in}$. This circumstance can give rise to an essential change of the superlattice ac response in conditions of the POP scattering suppression.

Making use Eq. (24) for the real part of the superlattice conductance $\text{Re}\,G(\omega, V_0)$ we obtain

$$\operatorname{Re} G(\omega, V_0) = G_0 \frac{1}{1+F_0^2} \frac{\left(1-F_0^2\right)\left[1-\delta(\omega\tau_{in})^2 + F_0^2\right] + (1+\delta)(\omega\tau_{in})^2}{\left[1-\delta(\omega\tau_{in})^2 + F_0^2\right]^2 + (1+\delta)^2 (\omega\tau_{in})^2}. \quad (31)$$

As can be seen from Eq. (31), the necessary condition for the terahertz gain in the superlattice $\operatorname{Re} G(\omega, V_0) < 0$ is determined by the following condition for the frequency $f = \omega/2\pi$:

$$f < f_{\max} = \frac{\Omega_B}{2\pi} \times \frac{1}{F_0} \sqrt{\frac{F_0^4 - 1}{F_0^2 + \delta^{-1}}}. \quad (32)$$

In the case of the absence of the applied dc electric field $F_0 = 0$ Eq. (31) yields the conventional Drude formula $\operatorname{Re} G(\omega, V_0 = 0) = G_0 / \left[1 + (\omega\tau_v)^2\right]$ describing the conductivity of the superlattice in thermal equilibrium. On the other hand, if the elastic scattering is neglected $r_{el} = 0$, we obtain the photon-assisted-tunneling result corresponding to Eq. (26) that can be presented as

$$\operatorname{Re} G(\omega, V_0) = G_0 \frac{1 + (\omega\tau_{in})^2 - F_0^2}{\left[1 + (F_0 + \omega\tau_{in})^2\right]\left[1 + (F_0 - \omega\tau_{in})^2\right]}. \quad (33)$$

In the high-frequency limit, i.e. at $\omega\tau_{in} \gg \max\{1, F_0\}$, Eq. (33) delivers the Drude-type roll-off of the ac conductance $\operatorname{Re} G(\omega, V_0) \propto G_0 / (\omega\tau_{in})^2$.

However, the situation becomes different in the presence of the elastic scattering. For example, if $r_{in} \ll r_{el}$, $\delta \ll 1$, $\omega \ll r_{el}$ Eq. (31) yields

$$\operatorname{Re} G(\omega, V_0) = G_0 \frac{1}{1+F_0^2} \frac{\left(1-F_0^2\right)\left(1+F_0^2\right) + (\omega\tau_{in})^2}{\left(1+F_0^2\right)^2 + (\omega\tau_{in})^2}. \quad (34)$$

In this case Eq. (34) delivers the frequency-independent value of the conductance $\operatorname{Re} G = G_0 / \left(1 + F_0^2\right)$ in the limit $\omega\tau_{in} \gg 1$. On the other hand, the value of the generalized second derivative $\overline{i}_0''(0, F_0)$ becomes $\overline{i}_0''(0, F_0) = -4\left(i_p/V_p^2\right) F_0 / \left(1 + F_0^2\right)^2$. This circumstance will result in a substantial change of the frequency dependence of the superlattice current responsivity comparatively to the case when the elastic scattering is ignored.

**D. Current responsivity**

In the present section we will study self-consistently the response of the superlattice to the intense terahertz fields, calculating the superlattice current responsivity $R_i$ which is defined as the ratio of the dc current change $\overline{\delta i_0}(\omega, V_0)$ to the incident terahertz power $P_i$[4, 5]

$$R_i(\omega, V_0) = \frac{\overline{\delta i_0}(\omega, V_0)}{P_i}. \tag{35}$$

Employing the equivalent circuit approach[4, 5] for the squared amplitude of the ac voltage across the superlattice $|V_1(\omega, V_0)|^2$ and for the current responsivity $R_i(\omega, V_0)$ for the case when the lateral size of the superlattice $D$ is much lager than its total superlattice thickness $L = N \times d$ we obtain

$$|V_1(\omega, V_0)|^2 = \frac{\gamma P_i \times 8 Z_A}{\left|1 + i Z_A \omega C_0 \varepsilon_{tot}(\omega, V_0) / \varepsilon_{SL}(0)\right|^2}, \tag{36}$$

$$R_i(\omega, V_0) = \frac{2\gamma \overline{i_0}''(\omega, V_0) \times Z_A}{\left|1 + i Z_A \omega C_0 \varepsilon_{tot}(\omega, V_0) / \varepsilon_{SL}(0)\right|^2}, \tag{37}$$

where $\gamma$ is the electromagnetic coupling factor between the antenna and the incident beam, $Z_A$ is the antenna impedance, $C_0 = \varepsilon_0 \varepsilon_{SL}(0) S / L$ is the static capacitance of the superlattice, $\varepsilon_{tot}(\omega, V_0) = \varepsilon_{SL}(\omega) + \sigma(\omega, V_0) / i\omega\varepsilon_0$ is the total dielectric function of the superlattice, $\varepsilon_{SL}(\omega) = (d_w + d_b) / \left[d_w \varepsilon_w^{-1}(\omega) + d_b \varepsilon_b^{-1}(\omega)\right]$ is the lattice dielectric function of the superlattice calculated using an effective medium theory[5], $\varepsilon_w(\omega)$ and $\varepsilon_b(\omega)$ are the lattice dielectric functions of the wells and barriers in the superlattice calculated by the Lorentzian equation[5] $\varepsilon_{w,b}(\omega) = \varepsilon_{w,b}(\infty) + \left[\varepsilon_{w,b}(0) - \varepsilon_{w,b}(\infty)\right] \omega_{TOw,b}^2 / \left(\omega_{TOw,b}^2 - \omega^2 + i\beta_{w,b}\omega\right)$, $\omega_{TOw,b}$ are the frequencies of the transversal polar-optical phonons in the wells and in the barriers, respectively, $d_w$ and $d_b$ are the wells and the barriers thicknesses, $\varepsilon_{w,b}(0)$ are the low-frequency lattice dielectric constants, $\varepsilon_{w,b}(\infty)$ are the lattice dielectric constants at the infinite frequency $\omega \to \infty$, $\beta_{w,b}$ are the optical-phonon damping constants, and $\sigma(\omega, V_0) = L G(\omega, V_0) / S$ is the superlattice complex conductance electrons conductivity.

Let us note that if the frequency of the incident terahertz fields $f = \omega / 2\pi$ is higher then the dielectric relaxation frequency in the superlattice, i.e. if $f \gg f_d = \sigma_0 / 2\pi\varepsilon(0)\varepsilon_0$, were $\sigma_0 = \sigma(\omega = 0, V_0 = 0)$ is the superlattice static conductivity at zero applied voltage, the contribution of the conduction electrons to the total dielectric function is negligible and, consequently, $\varepsilon_{tot}(\omega,V_0) \equiv \varepsilon_{SL}(\omega)$. In this case the frequency and the dc voltage dependencies of the responsivity $R_i(\omega,V_0)$ are determined by two factors. The first one, i.e. $1/|1+i\omega Z_A C_0|^2$, which does not depend on the applied dc voltage $V_0$, is responsible for the coupling efficiency between the antenna impedance $Z_A$ and the superlattice capacitive impedance $Z_0 = 1/i\omega C_0$. This factor determines the ZC-time roll-off frequency of the responsivity, i.e. the frequency

$$f_{Z_A C_0} = \frac{1}{2\pi Z_A C_0}. \tag{38}$$

The second factor is determined by the dc voltage and the frequency dependent generalized second derivative of the dc current-voltage curve $\bar{i}_0''(\omega,V_0)$. In particular, for the high frequencies of the incident waves, i.e. for $f > f_{Z_A C_0}$, $\omega\tau_{in} \gg 1$ the current responsivity $R_i(\omega,V_0)$ can be written as

$$R_i(\omega,V_0) = -4 \frac{i_p}{V_p^2 Z_A \omega^2 C_0^2} I(F_0) \times \left|\frac{\varepsilon_{SL}(0)}{\varepsilon_{SL}(\omega)}\right|^2 \times \text{Re}\frac{G(\omega,V_0)}{G_0}, \tag{39}$$

where $I(F_0) = 2F_0/(1+F_0^2)$ is the normalized dc current-voltage curve of the superlattice.

It is important to note that the current responsivity $R_i(\omega,V_0)$ described by Eq. (39) is directly proportional to the real part of the differential conductance $\text{Re}\,G(\omega,V_0)$. Consequently, the change in sign of the responsivity at some values of the applied dc voltage $V_0$, i.e. occurrence of the positive photocurrent $R_i(\omega,V_0) > 0$ should manifest about the presence of the terahertz gain in the superlattice at frequency $\omega$. On the other hand, the current responsivity $R_i(\omega,V_0)$ is directly proportional to the factor $i_p/V_p^2 \propto 1/r_{in}^{1/2}(r_{in}+r_{el})^{3/2}$ and, consequently, the suppression of the inelastic POP scattering $r_{in} \to 0$ should give rise to a dramatic enhancement of the responsivity value.

### E. DC voltage for terahertz gain

Using Eq. (31) for the value of the dc voltage $V_{NDC}$, above which the real part of the negative differential ac conductance of the superlattice becomes negative $\text{Re}\,G(\omega,V_0) < 0$, and the terahertz fields' gain takes place, we obtain

$$V_{NDC} = N \times \frac{\hbar}{e} \times \frac{\left\{\omega^2 + \left[\omega^4 + 4r_v^2\left(\omega^2 + r_{in}^2\right)\right]^{1/2}\right\}^{1/2}}{2^{1/2}}. \tag{40}$$

It is seen from Eq. (40) that in the limit $\omega \to 0$, the value of $V_{NDC}$ corresponds to that one derived from the dc current voltage curve of the superlattice $G(0,V_0) = di_0(V_0)/dV_0 < 0$, i.e. $V_{NDC} = V_p = N\hbar(r_{in}r_v)^{1/2}/e$. In the opposite case of the very high frequencies, i.e. at $\omega \gg r_v = r_{in} + r_{el}$, Eq. (40) results in the inverse Bloch oscillator condition[4, 20] for the positive photocurrent

$$V_{NDC} = N \times \frac{\hbar\omega}{e}, \tag{41}$$

which follows from the photon-assisted tunneling model described by Eq. (26). In terms of the dc electric field $E_0$ applied to the superlattice this resonance condition corresponds to the condition $\omega = \Omega_B$, where $\Omega_B = eE_0d/\hbar$ is the dc Bloch frequency of electrons.

However, in the intermediate case $r_{in} \ll \omega \ll r_v = r_{in} + r_{el}$ we obtain

$$V_{NDC} = N \times \frac{\hbar(\omega r_v)^{1/2}}{e} \gg N \times \frac{\hbar\omega}{e}. \tag{42}$$

This value of the $V_{NDC}$ does not correspond to any of the previous evaluations. This circumstance should essentially determine specifics of the frequency/dc voltage dependence of the superlattice current responsivity $R_i(\omega,V_0)$ in the terahertz-frequency band. On the other hand, the occurrence of the gain in superlattices at terahertz frequencies is often associated with the inverse Bloch oscillator effect[20]. However, Eq. (42) indicates that under conditions of the strong elastic scattering, the characteristic voltage $V_{NDC}$ may be considerably higher than the value resulting from Eq. (41).

## IV. RESULTS AND DISCUSSION

In some experiments describing the interaction of the intense terahertz fields with bulk superlattices[3] as well as in experiments devoted to investigation of the self-sustained current oscillation in the superlattices in the sub-terahertz frequency band[6-8] it has been observed that in agreement with Eq. (13) the peak current in the superlattice $i_p$ appears to be considerably less than the prediction of the Esaki-Tsu model described by Eq. (15). Actually it means that the elastic scattering can have a significant impact on the high-frequency response of the electrons even in the conventional bulk superlattices with no applied magnetic field. Of course, the role of the elastic scattering becomes much more substantial in the low-dimensional superlattices where the suppression of the POP scattering takes place.

To be specific in this section we analyze the superlattice response to strong terahertz fields for one of the samples described in Ref. 53. In that paper the drastic reduction of both the peak current $i_p$ and the peak voltage $V_p$ by strong magnetic fields have been experimentally observed. It was argued that the reduced dimensionality of electron motion caused by the magnetic field application leads to an increase of the inelastic scattering time and the corresponding decrease of the superlattice conductance electrons current.

The superlattice sample described in Ref. 53 had the well thickness $d_w = 7\ nm\ (GaAs)$, the barrier thickness $d_b = 1\ nm$, $(AlAs)$, the period $d = d_w + d_b = 8\ nm$, the allowed miniband width $\Delta = 20\ meV$, the forbidden minigap width $\Delta_G = 160\ meV$, the number of periods $N = 14$, the electron density $n = 3 \times 10^{16}\ cm^{-3}$. A $InAs$ layer with thickness of 0.8 monolayer was centered in each well. The measurements of the dc I-V curves were performed for the superlattice with the lateral size $D = 25\ \mu m$ at liquid helium temperatures $T = 4K$. In the absence of the magnetic field $B = 0$, by comparison of the measured values of the peak current $i_p$ and the peak voltage $V_p$ with the values given by Eqs. (17) and (18) the inelastic scattering time $\tau_{in} = 400\ fs$ and the elastic scattering time $\tau_{el} = 4\ fs$ have been found.

We note that in the present case both the thermal energy $kT\ (0.345\ meV)$ and the Fermi energy of degenerate electrons $\varepsilon_F\ (4.217\ meV)$ are much smaller than the superlattice miniband width, i.e. $kT, \varepsilon_F \ll \Delta$. This justifies neglecting of the mean electron energy in thermal equilibrium $\overline{\varepsilon}_0$ in Eqs. (6) and (7). We also point out that the width of the superlattice minigap in this case corresponds to the frequency $f_G = \Delta_G / 2\pi\hbar \approx 40\ THz$, so that the inter miniband transitions can be neglected in all terahertz frequency band $f \leq 40\ THz$.

*Figure 2* demonstrates the normalized peak current $i_p / i_{pET}$ calculated from Eq. (17) plotted as a function of the inelastic scattering time $\tau_{in}$. The inset in Fig. 2 shows the change of the dc current-voltage curves for $\tau_{el} = 4\ fs$ and for the inelastic scattering relaxation time values $\tau_{in}$ varying from $\tau_{in} = 400\ fs$ up to $\tau_{in} = 6000\ fs$. The calculations are in a good agreement with the observations of Ref. 53 that corresponds to the magnetic field change from $B = 0\ T$ up to $B = 23\ T$. Figure 2 also shows that in this range of magnetic fields $B$ (inelastic scattering relaxation time values $\tau_{in}$) the normalized current as a function of $\tau_{in}$ obeys to the power law $i_p / i_{pET} = (\tau_{el} / \tau_{in})^{1/2}$ observed in Ref.53. Therefore, the found values of the inelastic scattering relaxation time values $\tau_{in}$ as a function of the magnetic field amplitude $B$ can be satisfactorily used for our study of the terahertz response of the superlattice.

As can be seen from the inset of Fig. 2, in a full agreement with the experimental data presented in Ref. 53, increasing of the inelastic scattering relaxation time $\tau_{in}$ does not change the low-field superlattice conductance $G(\omega \to 0, V_0 \to 0) = dI_0(V_0 \to 0)/dV_0$. This circumstance is easy to understand if one considers that when the inelastic scattering rate is negligible $r_{in} \ll r_{el}$, according to Eq. 20 the superlattice electron mobility $\mu_0$ in low fields is determined solely by the elastic scattering rate $r_{el}$ which remains unchanged upon the application of the strong magnetic fields. In this case the low-field superlattice conductance $G(\omega \to 0, V_0 \to 0) = dI_0(V_0 \to 0)/dV_0 = G_0$ remains unchanged upon the application of the strong magnetic field $B$ as well.

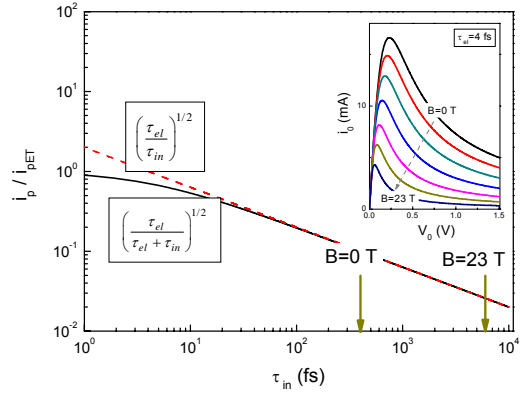

FIG. 2. The normalized peak current $i_p / i_{pET}$ in the superlattice plotted as a function of the inelastic scattering relaxation time $\tau_{in}$ for the fixed value of the elastic scattering time $\tau_{el} = 4\,fs$. The inset shows the change of the dc current-voltage curves of the superlattice when the applied longitudinal magnetic field is varied from $B = 0\,T$ up to $B = 23\,T$.

According to Eq. (37), the value and the frequency dependence of the current responsivity of the superlattice $|R_i(\omega, V_0)|$ is essentially dependent on the high-frequency conductivity $\sigma(\omega, V_0)$. On the other hand, the maximum frequency of the terahertz fields' gain in the superlattice $f_{max}$ is also determined by the frequency dependence of the

conductivity $\text{Re}\left[\sigma(\omega,V_0)\right]$. Therefore both of these fundamental characteristics of the superlattice terahertz response may be strongly influenced by the presence of the elastic scattering of electrons.

*Figure 3* shows the dependence of the frequency $f_{max}$ on the dc electric field strength $E_0$ for $r_{in} = 2.5 \times 10^{12}\, s^{-1}$ and for the different values of the elastic scattering rates $r_{el}$ varying from $r_{el} = 0$ to $r_{el} = 2.5 \times 10^{14}\, s^{-1}$.

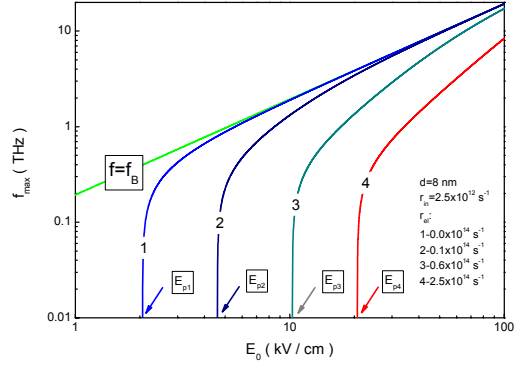

FIG. 3. Maximum frequency of the negative differential conductivity (NDC) effect in a superlattice $f_{max}$ plotted as a function of the dc electric field strength $E_0$ for different values of the elastic scattering rates $r_{el}$.

It is seen from Fig. 3 the ac NDC effect in the superlattice takes place only when the applied dc electric field $E_0$ exceeds the value of the peak dc electric field $E_p = \hbar\left[r_{in}(r_{in} + r_{el})\right]^{1/2}/ed$. The values of the peak dc field $E_p$ calculated for corresponding values of $r_{el}$ are indicated in Fig. 3 by arrows. The peak dc electric field $E_p$ calculated for $r_{in} = 2.5 \times 10^{12}\, s^{-1}$ and $r_{el} = 2.5 \times 10^{14}\, s^{-1}$ ($\tau_{in} = 400\, fs$, $\tau_{el} = 4\, fs$) is found to be in a good agreement with the experimental observations reported in Ref. 53 in the case of zero magnetic fields $B = 0\, T$. Dependence of the frequency $f_{max}$ on the dc electric field $E_0$ for this case is presented in Fig. 3 by the curve 4 (red line). Importantly, that if the applied dc electric field $E_0$ exceeds the peak field $E_p$ the frequency $f_{max}$ is considerably less then the intuitively supposed estimate $f_{max} \approx f_B = eE_0 d/2\pi\hbar$. This conclusion is in line with the results of numerical calculations taking into account the presence of the elastic scattering by the Monte Carlo method[62].

*Figure 4* demonstrates the dc voltage $V_{NDC}$ above which the negative differential conductance of the superlattice $\text{Re}\,G(\omega,V_0)$ becomes negative, i.e. the characteristic frequency of the terahertz field gain, plotted as a function of the ac field frequency $f = \omega/2\pi$ for different values of the inelastic scattering rate $r_i$. It is seen that at almost zero frequencies, the dc voltage $V_{NDC}$ tends to the value of the peak voltage $V_p$ described by Eq. (18), i.e. $V_{NDC} \to V_p$ at $f \to 0$. And, therefore, the dc voltage $V_{NDC}$ decreases with decreasing of the inelastic scattering rate $r_i$.

In the high-frequency band, i.e. for $f \geq 1\ THz$, the frequency dependence of the dc voltage $V_{NDC}$ is well described by Eq. (42). In contrast to the low-frequency band ($f < 1\ THz$), in this case the dc voltage $V_{NDC}$ considerably increases with increasing of the ac field frequency $f$. It is important to note that in our case of the strong elastic scattering the characteristic voltage $V_{NDC}$ considerably exceeds the values obtained based on the Eq. (41) describing the effect of the inverse Bloch oscillator.

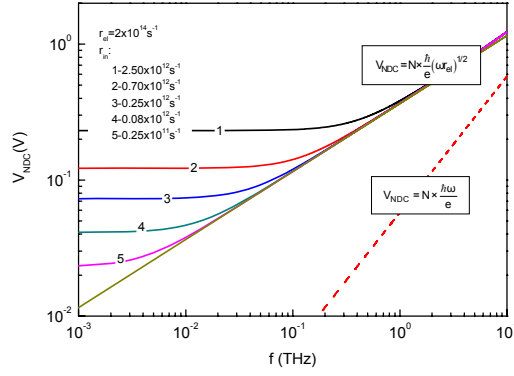

FIG. 4. The dc voltage $V_{NDC}$ above which the negative differential conductance of the superlattice $\operatorname{Re} G(\omega, V_0)$ becomes negative (the region of the terahertz field gain) plotted as a function of the ac field frequency $f = \omega/2\pi$ for different values of the inelastic scattering rate $r_i$. The superlattice parameters are taken from Ref. 53.

*Figure 5* shows the results of calculations of the absolute value of the current responsivity $|R_i(\omega, V_0)|$ as a function of the frequency $f = \omega/2\pi$ at different values of the superlattice lateral size $D$ performed based on Eq. (37). We assume the dc voltage $V_0 = V_p$, the value of the corner-cube antenna impedance $Z_A = 150\ \Omega$, and the electromagnetic coupling factor $\gamma = 0.5$. It is seen from Fig. 4 that at frequencies $f \geq f_{Z_A C_0} = 1/2\pi Z_A C_0$, the roll-off of the current responsivity $|R_i(\omega, V_0)|$ caused by the mismatch between the antenna impedance and the impedance of the superlattice takes place.

We note that all terahertz frequency band $f = \omega/2\pi = (1-10)$ $THz$ ranges within the inelastic scattering frequency $f_{in} = r_{in}/2\pi = 0.40$ $THz$ and the elastic scattering frequency $f_{el} = r_{el}/2\pi = 39.8$ $THz$. This frequency band includes the region of the polar optical phonon frequencies, i.e. $f_{TOw} = \omega_{TOw}/2\pi = 8.01$ $THz$ and $f_{TOb} = \omega_{TOb}/2\pi = 10.8$ $THz$ for the wells and barriers, respectively. As it can be seen from Fig. 5, the huge resonant futures in the frequency dependence of $|R_i(\omega,V_0)|$ take place due to effects of screening and "anti-screening" of the terahertz fields[5] at frequencies $f \approx f_{TOw}$ and $f \approx f_{LOw}$, where $f_{LOw,b} = f_{TOw,b} \times \sqrt{\varepsilon(0)/\varepsilon(\infty)}$ is the longitudinal polar-optical phonon frequency, and $f_{TOw}$ is the transversal polar-optical phonon frequency in the wells in barriers.

Let us note that at $f > f_{in} = r_{in}/2\pi$ the responsivity $|R_i(\omega,V_0)|$ follows rather slow decay law, i.e. $|R_i(\omega,V_0)| \propto 1/f^2$, changing to the more steep decay, i.e. to $|R_i(\omega,V_0)| \propto 1/f^4$, at $f > f_{el} = r_{el}/2\pi$. Such frequency behavior of the responsivity is significantly different from the results obtained within the framework of the photon-assisted tunneling theory[5] ($r_{el} = 0$), where $|R_i(\omega,V_0)| \propto 1/f^4$ at frequencies $f > f_{in} = r_{in}/2\pi$. This substantial difference arises due to the fact that in the frequency range $f_{in} < f < f_{el}$ according to Eq. (34) the real part of the superlattice conductivity $\text{Re}\,G(\omega,V_0)$ is independent of the frequency $f = \omega/2\pi$.

It is also important that in the terahertz frequency region $f = (1-10)$ $THz$ the absolute value of the current responsivity $|R_i(\omega,V_0)|$ is increasing very rapidly with decreasing lateral size of the superlattice $D$. For example, at frequency $f = 4$ $THz$ the decrease of the superlattice lateral size $D$ from 25 $\mu m$ to 1 $\mu m$ leads to the increase of the responsivity by more than three orders of magnitude. In connection with this, we will analyze the influence of the POP scattering suppression on the high-frequency response of the superlattices for the samples with rather low lateral size.

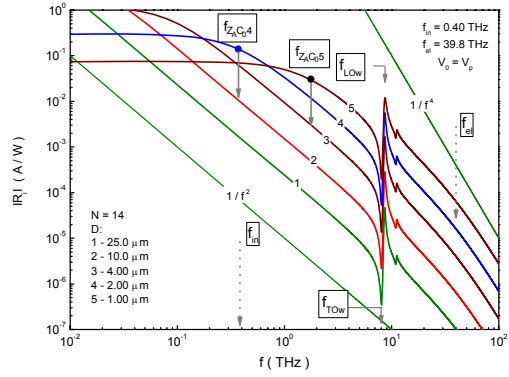

FIG. 5 Absolute value of the superlattice current responsivity $\left|R_i\left(\omega,V_0\right)\right|$ plotted as a function of the incident terahertz field frequency $f = \omega/2\pi$ for different values of the superlattice lateral size $D$ at $V_0 = V_p$. The superlattice parameters are taken from Ref. 53.

*Figure 6* demonstrates the results of the calculation of the current responsivity $\left|R_i\left(\omega,V_0\right)\right|$ for the superlattice sample parameters experimentally investigated in Ref. 53. Firstly, in Fig. 6(a) we show the dc current-voltage curves of the superlattice with the lateral size $D = 25$ $\mu m$, and the elastic scattering rate $r_{el} = 2.5\times10^{14}$ $s^{-1}$ for diminishing values of the inelastic scattering rates ranging from $r_{in} = 2.5\times10^{12}$ $s^{-1}$ to $r_{in} = 0.25\times10^{12}$ $s^{-1}$. It is seen that both the value of the peak current $i_p \propto \sqrt{r_{in}/(r_{in}+r_{el})}$ and the peak voltage $V_p \propto \sqrt{r_{in}(r_{in}+r_{el})}$ are dropping down with decreasing of the inelastic scattering rate $r_{in}$. The interval between these values corresponds to the range found experimentally in Ref. 53 when the strong magnetic field $B = (0-23)$ $T$ (suppressing the polar-optical phonon scattering) was applied to the sample.

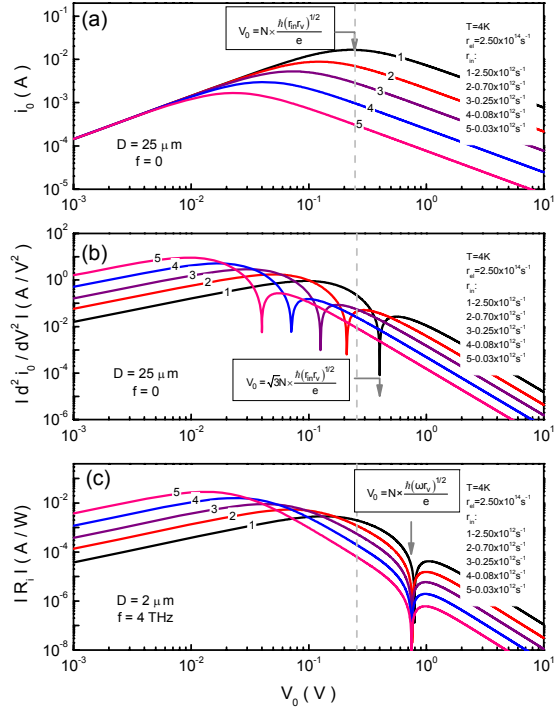

FIG. 6. (a) The dc current-voltage curves $i_0(V_0)$ as a function of the applied dc voltage $V_0$ calculated for different values of the inelastic scattering rates $r_{in}$ corresponding to experiments described in Ref. 53. (b) The second derivative of the dc current-voltage curves $\left| d^2 i_0(V_0)/dV_0^2 \right|$ as a function of the applied dc voltage $V_0$ calculated for the same values of the superlattice parameters. (c) The absolute value of the current responsivity $\left| R_i(\omega, V_0) \right|$ as a function of the applied dc voltage $V_0$ calculated for the same values of the superlattice parameters and for the reduce value of the superlattice lateral size $D = 2 \ \mu m$ at frequency $f = 4 \ THz$.

At low frequencies of the incident ac fields, i.e. at $\omega \ll r_{in}$, according to Eq. (37) the value of the current responsivity $R_i(\omega, V_0)$ is proportional to the ordinary second derivative of the dc current-voltage curve of the superlattice. In Fig. 5 (b) we show the absolute value of the ordinary second derivative of the dc current-voltage curve $\left| d^2 i_0(V_0)/dV_0^2 \right|$ as a function of the dc voltage $V_0$. It is seen that at rather high dc voltage $V_0$, i.e. at $V_0 \geq 0.2 \ V$, the second derivative diminishes with increasing of the inelastic scattering rate $r_i$. In contrast, at low voltage $V_0 < 0.2 \ V$ the decrease in value of $r_i$ gives rise to a giant (greater than two orders of magnitude) increase in the absolute value of the second derivative $\left| d^2 i_0(V_0)/dV_0^2 \right|$. The dips due to the change of sign of the second derivative $d^2 i_0(V_0)/dV_0^2$ also takes place at $V_0 = \sqrt{3} V_p = \sqrt{3} N \times \hbar (r_{in} r_v)^{1/2} / e$.

In Fig. 5 (c) we show the absolute value of the current responsivity $|R_i(\omega, V_0)|$ calculated from Eq. 37 as a function of the applied dc voltage $V_0$ for the same values of the inelastic scattering rate $r_{in}$ assuming the reduced value of the superlattice lateral size $D = 2~\mu m$ at the frequency $f = 4~THz$. In this case the condition for the incident terahertz frequency field $r_{in} \ll \omega \ll r_v = r_{in} + r_{el}$ is well satisfied. As in the previous case at rather high dc voltage $V_0$, i.e. at $V_0 \geq 0.2~V$, the absolute value of the current responsivity $|R_i(\omega, V_0)|$ diminishes with increasing of the inelastic scattering rate $r_i$. In contrast, at low voltage $V_0 < 0.2~V$ decrease in the value of $r_i$ gives rise *to a giant (greater than two orders of magnitude) enhancement of the absolute value of the current responsivity* $|R_i(\omega, V_0)|$. On the other hand, in this case the voltage positions of the dips at $V_0 = N \times \hbar (\omega r_v)^{1/2} / e$ in the dependence of $|R_i(\omega, V_0)|$ on the dc bias voltage $V_0$ very weakly depend on the inelastic scattering rate $r_i$.

Qualitatively, specifics of the current responsivity enhancement in condition of the polar-optical phonon scattering suppression can be easily understood if one considers that even at high frequencies $\omega \tau_{in} \gg 1$ absorption of the incident terahertz fields is still efficient due to the presence of the elastic scattering of electrons. At the same time, reducing the rates of inelastic scattering leads to the more efficient electronic heating. This, in turn, gives rise to the decrease of the electron mobility and, therefore, to a well pronounced electronic bolometric effect.

It is important to note that in Ref. 53 at very strong magnetic fields $B > 21~T$ corresponding to the condition $\hbar \omega_c > \hbar \omega_{LO} = 36~meV$, where $\hbar \omega_{LO}$ is the energy of the LO phonons in GaAs, some features related to a stark magneto-phonon resonance[39-41] have been observed. Their voltage position was insensitive to $B$ and was given by the conditions $\hbar \omega_B = \hbar \omega_{LO}/2$ and $\hbar \omega_B = \hbar \omega_{LO}$ that corresponds to the dc negative differential conductance region, i.e. to the condition $V_0 > V_p = N \hbar (r_{in} r_v)^{1/2} / e$. However, it should be noted that in our case the responsivity enhancement takes place in the region of the dc bias voltages lying below the threshold voltages $V_0 < V_p = N \times \hbar (r_{in} r_v)^{1/2} / e$, i.e. in the region where the superlattice dc differential conductance is positive for the each given magnetic field amplitude $B$. Therefore, the space charge wave instability in the superlattice[6-8], as well as the stark magneto-phonon resonance[39-41] does not interfere with the terahertz photon detection.

Let us finally note that an experimental investigation of the non-linear response of semiconductor superlattices in strong magnetic fields could have been performed with such instruments as free-electron lasers[66]. Recently, the free-electron laser[66] has been used for direct determination of the electron effective mass in GaAsN non-parabolic bands by means of the terahertz cyclotron resonance spectroscopy in magnetic fields up to 60 T.

## V. CONCLUSIONS

……In conclusion, we investigated the high-frequency response of the low-dimensional semiconductor superlattices in condition of a suppression of the inelastic scattering of electrons caused by polar-optical phonons. For solution of this problem we developed a balance equations approach describing the time evolution of the electrons mean velocity and the mean electron energy in strong ac/dc electric fields. This approach takes into account both the elastic and the inelastic scattering of electrons. Our treatment allows one to consider the influence of the Bloch dynamic of electrons in a superlattice miniband side by side with the influence of an electron heating on the superlattice current responsivity in the terahertz frequency band.

Based on this approach we calculated the dc current-voltage curve of the superlattice investigated experimentally when the inelastic scattering of electrons caused by the polar-optical phonons was suppressed by the application of a strong magnetic field. Using the analytical expression for the dc current voltage curves of the superlattice we estimated the change of the inelastic scattering rate caused by the strong magnetic fields.

We obtained analytical expressions for the ac conductance and for the dc current change in the superlattice irradiated with strong terahertz fields. We found that both of them can not be described properly by photon-assistant tunneling equations if the elastic scattering of electrons is fairly significant. We showed that both the frequency dependence of the superlattice conductance and the maximum frequency of the terahertz gain can be very different from predictions of the photon-assistant tunneling theory.

We derived an analytical expression for the superlattice current responsivity in the terahertz frequency band taking into account the influence of the resonant interaction of the terahertz fields with the polar-optical phonons. We demonstrated that at rather high frequencies the change of the responsivity sign (the transition from the negative values of the photocurrent to the positive ones) corresponds to the transition from loss to gain even in the presence of the strong elastic scattering.

We studied the maximum frequency of the terahertz fields gain in the presence of the elastic scattering in strong applied dc electric fields. We found that the presence of the elastic scattering can dramatically reduce the value of the maximum frequency of the terahertz fields gain in comparison with the value of the Bloch frequency of electrons. Our investigation showed that the reduction of the superlattice lateral size can provide a dramatic enhancement of the current responsivity magnitude in the terahertz frequency band.

Our study shows that the suppression of the inelastic scattering caused either by the reduction of the superlattice dimensionality or by the strong magnetic field application can give rise to a huge enhancement of the current responsivity of the superlattice in the terahertz frequency band. This enhancement can be interpreted in terms of the well pronounced bolometric effect occurring due to the effective electron heating in the superlattice by the incident terahertz fields.

## VI. ACKNOWLEDGMENTS

The support from the Program for the Basic Research of the Russian Academy of Sciences is gratefully acknowledged. This work was supported in part by the Ministry of Education and Science of the Russian Federation.